\documentclass[10pt,twoside]{Lowx2011}
\usepackage{epsf,amsmath}
\usepackage{graphicx}

\newcommand{\pom}{\tt I\! P}
\begin{document}

\title{EXCLUSIVE GLUEBALL PRODUCTION AT THE LHC
\thanks{This work is
supported by the funding agency CNPq, Brazil}}

\author{\underline{M. V. T. MACHADO} and M. L. L. da SILVA\\ \\
High Energy Physics Phenomenology Group, GFPAE  IF-UFRGS\\
Caixa Postal 15051, CEP 91501-970, Porto Alegre, RS, Brazil\\
E-mail: magnus@if.ufrgs.br}

\maketitle

\begin{abstract}
\noindent In this contribution we summarize recent results on the computation of cross sections for glueball candidates production in quasi-real photon-photon collisions and on central diffraction processes, i.e. double Pomeron exchange, in heavy ion interactions at the LHC are computed. In particular, we provide predictions for the production of exotic mesons $f_0(1500)$, $f_0(1710)$ and $X(1835)$. The rates for these distinct production channels are compared.
\end{abstract}



\markboth{\large \sl \hspace*{0.25cm}\underline{M. V. T. Machado} \& M. L. L. da Silva
\hspace*{0.25cm} Low-$x$ Meeting 2011} {\large \sl \hspace*{0.25cm} EXCLUSIVE GLUEBALL PRODUCTION AT THE LHC}

\section{Introduction}

The gluon self-coupling in QCD opens the possibility of existing bound states of pure gauge fields known as glueballs ($G$), which  are predicted by several theoretical formalisms and by lattice calculations \cite{Vento,Crede}.  Many mesons  have stood up as good candidates for the lightest glueball in the spectrum and in particular the scalar sector ($J^{PC} = 0^{++}$) seems promising. The mesons $f_0(1500)$ and the $f_0(1710)$ have been considered  the principal candidates for the
scalar glueball\cite{aleph,closekirk}. However, in
this mass region the glueball state will mix strongly with nearby $q\bar{q}$ states \cite{closekirk,amsler}. More recently, the BES collaboration observed a new resonance called $X(1835)$ \cite{bes}.

Recently, the clean topologies of exclusive particle production in electromagnetic interactions hadron-hadron
and nucleus-nucleus collisions mediated by colorless exchanges such the QCD Pomeron or two photons have attracted
an increasing interest \cite{upcs}. The cross sections for these processes are smaller than the correspondent
inclusive production channels, which it is compensated by a more favorable signal/background relation.
Experimentally, exclusive events are identified by  large rapidity gaps on both sides of the produced central
system and the survival of both initial state particles scattered at very forward angles with respect to the beam.

In this contribution, we summarize some results presented in Ref. \cite{Magno_PRC} on exclusive glueball production in two-photon and Pomeron-Pomeron interactions in coherent
nucleus-nucleus collisions at the LHC. We also present new results for the proton-proton collisions at 7 TeV. In the two-photon channel, the photon flux scales as the
square charge of the beam, $Z^2$, and then the corresponding cross section is highly enhanced by a factor
$\propto Z^4\approx 10^7$ for gold or lead nuclei. A competing channel, which produces similar final state
configuration, is the central diffraction  (CD) process. Such a reaction is modeled in general by two-Pomeron
interaction.  One goal of the present investigation is to
compare the cross sections for these two channels in the production of glueball candidates.

\section{Theoretical framework and phenomenological investigations}

Let us start with the glueball candidate production in photon-photon scattering at coherent heavy ion collisions using the
Weizs\"acker - Williams approximation (EPA approximation). In such an approach, the cross section for a two
quasi-real photon process to produce a glueball state, $G$, at center-of-mass energy $W_{\gamma \gamma}$ factorises
into the product of the elementary cross section for $\gamma \gamma \rightarrow G$ convoluted with the equivalent
photon spectra from the colliding ions \cite{upcs}:
\begin{eqnarray}
\sigma_{\gamma\gamma} (AA \rightarrow A+G+A) = \int \frac{dk_1}{k_1} \frac{dk_2}{k_2} \frac{dn_{\gamma}}{dk_1}
\frac{dn_{\gamma}}{dk_2}\sigma\,(\gamma\gamma\rightarrow G), \label{sigfoton}
\end{eqnarray}
where $k_{1,2}$ are the photon energies and $dn/dk$ is the photon flux at the energy $k$ emmited by the hadron $A$.
The photon energies determine the center-of-mass energy $W_{\gamma \gamma}= \sqrt{4k_1k_2}$ and the rapidity $Y$ of
the produced system. Namely, one has $k_{1,2}=(W_{\gamma\gamma}/2)\exp (\pm Y )$ and $Y=(1/2)\ln \,(k_1/k_2)$.  In
addition, $\sqrt{s_{NN}}$ is the center-of-mass energy of the ion-ion system and the Lorentz relativistic factor  is
given by $\gamma_L= \sqrt{s_{NN}}/(2m_N)$. In the numerical calculations we use
$\sqrt{s_{NN}}=5.5$ TeV  and $\gamma_L= 2930$ for the LHC. We also will present prediction for the ALICE run at $\sqrt{s}=2.76$ TeV (see Table 1)

In the EPA approximation, the flux of equivalent photons from a relativistic particle of charge $Z$ is determined
from the Fourier transform of its electromagnetic field. For an extended charge with electromagnetic form factor,
$F_A(Q^2)$, the energy spectrum can be computed as,
\begin{eqnarray}
\frac{dn_{\gamma/A}\,(x)}{dk} = \frac{\alpha\,Z^2}{\pi}\,\frac{A(x)}{x}\int\frac{Q^2-Q_{\mathrm{min}}^2}{Q^4}
\,|F_A(Q^2)|^2\,dQ^2,
\label{fotflux}
\end{eqnarray}
where $x=k/E$ is the fraction of the beam energy carried by the photon and $A(x)=1-x+(1/2x^2)$. Moreover,
$\alpha=1/137$ and $Q^2$ is the four-momentum transfer squared from the charge, with
$Q_{\mathrm{min}}^2\approx (x m_N)^2/(1-x)$.

The glueball production in two-photon fusion can be calculated using the narrow resonance approximation \cite{BKT}:
\begin{eqnarray}
\sigma\,(\gamma\gamma\rightarrow G) = (2J+1)\,\frac{8\pi^2}{M_G}\,\Gamma(G\rightarrow \gamma \gamma)\,\delta
\left(W_{\gamma\gamma}^2 -  M_G^2  \right),
\label{twophotres}
\end{eqnarray}
where $\Gamma(G\rightarrow \gamma \gamma)$ is the partial two-photon decay width of $G$, $M_G$ is the glueball mass
and $J$ is the spin of the state $G$. Here, we compute the production rates for the mesons $f_{0}(1500)$,
$f_{0}(1710)$ and $X(1835)$ \cite{PDG}, respectively. The reason is due to they  have been mentioned as possible glueball candidates by phenomenologists \cite{Vento,Crede}.

The predictions for the two-photon component are in practice somewhat difficult as the branching ratios have not been measured. To compute numerical values for the meson (glueball) cross section in two-photon reactions estimates for the two-photon decay widths are needed. The determination of them depend upon whether the meson state is a pure quarkonium, pure gluonic or a mixed hybrid state. A detailed discussion  for the main fetures in each case can be found in Ref. \cite{Magno_PRC}, where it was verified that the pure glueball production gives the lowest cross section. Accordingly, the estimates for the two-photon width for a pure glueball resonance are $\Gamma(f_0(1500) \to \gamma \gamma) \simeq 0.77$ eV, $\Gamma(f_0(1710) \to \gamma \gamma) \simeq 7.03$ eV  and $\Gamma(X(1835) \to \gamma \gamma) \simeq 0.021$ keV. The widths are about three orders of magnitude smaller that for pure $q\bar{q}$ states.  Therefore, as the two-photon cross section scales as $(2J+1)\Gamma \,(R\rightarrow \gamma\gamma)$, Eq. (\ref{twophotres}), one can consider the experimental feasibility of using peripheral heavy-ion collisions to determine the nature of the resonances discussed above.

Let us now focus on the Pomeron-Pomeron channel. In particular, we consider  the central diffraction (double Pomeron
exchange, DPE) in nucleus-nucleus interactions. As a starting point we compute the DPE proton-proton cross section
making use of the  Bialas-Landshoff  \cite{Land-Nacht,Bial-Land} approach. It is believed that this non-perturbative approach is a reasonable choice due to the light mass of glueballs candidates considered and we are interested in the central exclusive
 DPE production of glueball states. In the exclusive DPE event the central
object $G$ is produced alone, separated from the outgoing hadrons by rapidity gaps,
$pp\rightarrow p+\text{gap}+G+\text{gap}+p$.  In approach we are going to use, Pomeron exchange corresponds to the exchange of a
pair of non-perturbative gluons which takes place between a pair of colliding quarks. The scattering matrix is given by,
\begin{eqnarray}
\mathcal{M}  =  \mathcal{M}_{0}\left(  \frac{s}{s_{1}}\right)  ^{\alpha(t_{2})-1}\left(  \frac{s}{s_{2}}\right)
^{\alpha(t_{1})-1}\,F(t_{1})\,F(t_{2}) \exp\left(  \beta\left(  t_{1}+t_{2}\right)  \right)\,  S_{\text{gap}}\left(\sqrt{s} \right).
\label{M_all}
\end{eqnarray}
Here $\mathcal{M}_{0}$ is the amplitude in the forward scattering limit ($t_1=t_2=0$). The standard Pomeron Regge
trajectory is given by $\alpha\left(  t\right)=1+\epsilon+\alpha^{\prime}t$ with
$\epsilon\approx 0.08,$ $\alpha^{\prime}=0.25$ GeV$^{-2}$. The momenta of incoming (outgoing) protons are labeled by $p_1$
and $p_2$ ($k_1$ and $k_2$), whereas the glueball momentum is denoted by $P$. Thus, we can define the following quantities
appearing in Eq. (\ref{M_all}): $s=(p_{1}+p_{2})^{2}$, $s_{1}=(k_{1}+P)^{2},$ $s_{2}=(k_{2}+P)^{2},$
$t_{1}=(p_{1}-k_{1})^{2}$ , $t_{2}=(p_{2}-k_{2})^{2}$. The nucleon form-factor is given by $F_p\left(  t\right)  $ = $\exp(b
t)$ with $b=$ $2$ GeV$^{-2}$. The phenomenological factor $\exp\left(  \beta\left(  t_{1}+t_{2}\right)  \right)
$ with $\beta$ $=$ $1$ GeV$^{-2}$ takes into account the effect of the momentum transfer dependence of the non-perturbative
gluon propagator. The factor $S_{\text{gap}}$ takes the gap survival effect into account $i.e.$ the probability
($S_{\text{gap}}^{2}$) of the gaps not to be populated by secondaries produced in the soft rescattering.  For our purpose
here, we will consider $S_{\mathrm{gap}}^2=0.032$ at $\sqrt{s}=5.5$ TeV in nucleon-nucleon collisions.

Following \cite{Bial-Land} we find
$\mathcal{M}_{0}$ for colliding hadrons,
\begin{eqnarray}
\mathcal{M}_{0}=32 \,\alpha_0^2\,D_{0}^{3}\,\int d^{2}\vec{\kappa}\,p_{1}^{\lambda}V_{\lambda\nu}^{J}p_{2}^{\nu}\,
\exp(-3\,\vec{\kappa }^{2}/\tau^{2}),
\label{M_o}
\end{eqnarray}
where $\kappa $ is the transverse momentum carried by each of the three gluons. $V_{\lambda\nu}^{J}$ is the
$gg\rightarrow G^{J}$ vertex depending on the polarization $J$ of the $G^{J}$ glueball meson state. For the
cases considered here, $J=0$, one obtains the following result
\cite{Bial-Land,KMRS}:
\begin{equation}
p_{1}^{\lambda}V_{\lambda\nu}^{0}p_{2}^{\nu}=\frac{s\,\vec{\kappa}^{2}
}{2M_{G^{0}}^{2}}A, \label{p_1Vp_2}%
\end{equation}
where $A$ is expressed by the mass $M_{G}$ and the width $\Gamma (gg\rightarrow G)$ of the glueball meson through the relation:
\begin{equation}
A^{2}= 8\pi M_G\,\Gamma (gg\rightarrow G). \label{A^2}
\end{equation}
For obtaining the two-gluon decays widths the following relation is used,
$\Gamma \,(G\rightarrow gg)=\mathrm{Br}\,(G\rightarrow gg)\,\Gamma_{tot}(G)$. Here, we will be conservative and assume the resonance to be a pure glueball. This fact translates into an upper bound for the exclusive DPE production as the cross section scales with $\Gamma\,(R\rightarrow gg)$. Following Ref. \cite{close}, the two-gluon width can be computed from the resonance branching fraction in $J/\psi$ radiative decay, $\mathrm{Br}\,(\psi\rightarrow \gamma \,G)$. For the candidates of interest here one obtains:
\begin{eqnarray}
\mathrm{Br}\,(G(J^{PC})\rightarrow gg) = \frac{8\pi(\pi^2-9)\,Br[\psi\rightarrow \gamma\,G(J^{PC})]}{c_R\,x|H_J(x)|^2\,\Gamma_{tot}}\frac{M_{\psi}^2}{M_G},
\end{eqnarray}
where the function $H_J((x)$ is determined in the non-relativistic quark model (NRQM) (see appendix of Ref. \cite{close}) and $c_R$ is a numerical constant ($C_R=1, \,2/3,\,5/2$ for $J^{PC}=0^{-+},\,0^{++},\,2^{++}$, respectively). The masses of $J/\psi$ and of resonance are $M_{\psi}$ and $M_G$, respectively,  and $x=1-(M_G^2/M_{\psi}^2)$. Based on equations above, in Ref. \cite{close} the following values for the branching fractions for scalar glueballs candidates are obtained:  $\mathrm{Br}\,[f_0(1500)]=0.64\pm 0.11$, $\mathrm{Br}\,[f_0(1710)]=0.52\pm 0.07$. For the pseudoscalar $X$ the situation is less clear due to small information on its decaying channels in radiative $J/\psi$ decays. Here, we set the limit case $\mathrm{Br}\,[X(1835)]=1$ and notice that the branching would be about 30 \% smaller. As discussed in \cite{Magno_PRC}, a consequence on the small deviation for the branching fraction in pure $q\bar{q}$ and glueball resonance is the difficulty in testing their nature using the exclusive diffractive data.

In order to calculate the $AA$ cross section the procedure presented in Ref. \cite{Pajares} is considered, where the central
diffraction and single diffraction cross sections in nucleus-nucleus collisions are computed using the so-called
{\it criterion C} (we quote Ref. \cite{Pajares} for further details).  Using the profile function for two colliding nuclei, $T_{AB}=\int d^2\bar{b}\,T_A(\bar{b})\,T_B(b-\bar{b})$, the final expression for CD  cross section in $AA$
collisions is given by \cite{Pajares}:
\begin{eqnarray}
\sigma^{\mathrm{CD}}_{AA} =  A^2\int d^2b \,T_{AA}(b)\,\exp\left[-A^2\,\sigma^{in}_{pp}\,T_{AA}(b)\right] \,\sigma^{\mathrm{CD}}_{pp}.
\label{sdxs1}
\end{eqnarray}
where $\sigma^{in}_{pp}$ and $\sigma^{\mathrm{CD}}_{pp}=S_{\mathrm{gap}}^2\times\sigma^{\mathrm{CD}}_{pp}(\sqrt{s}) $ are the inelastic and CD cross sections in proton-proton case, respectively. Using Woods-Saxon nuclear densities and considering the inelastic cross section $\sigma^{in}_{pp}=73$ mb for LHC energy, $\sqrt{s_{AA}}=5.5$ TeV, we compute the CD cross section for nuclear collisions. The values for the inelastic cross section are obtained from $\mathrm{DPMJET}$ \cite{DPM}, where the scattering amplitude is parameterized using $\sigma_{tot}$, $\rho$ and elastic slope (these parameters are taken as fitted by the $\mathrm{PHOJET}$ model \cite{Engel}). We notice that for LHC energy the effective
atomic number dependence is proportional to $A^{1/3}$, which means that the nuclear CD cross section is only one order of magnitude larger than the nucleon-nucleon cross section.

In what follows,  we compare the two production channels and investigate the main theoretical uncertainties. We provide estimates of
cross sections and event rates for both processes the LHC energies at the heavy ion and proton-proton mode.

\section{Results and discussions}

In Table 1 numerical results for the two-photon and Pomeron-Pomeron processes are presented and discussed. The
cross sections for glueball candidates production in photon-photon fusion at energy of $\sqrt{s}=5.5$ TeV for lead-lead collisions are shown.  The cross sections are sufficiently large for experimental measurement. The event rates can be obtained using the beam luminosity \cite{upcs}: for LHC one has ${\cal L}_{\mathrm{PbPb}} = 5 \cdot 10^{26}$ cm$^{-2}$s$^{-1}$, which
produces the following number of events. One has  $3.6\cdot 10^2$, $2\cdot 10^3$ and $4$ for $f_0(1500)$, $f_0(1710)$ and
$X(1835)$, respectively,  in the nominal LHC running time with ions of $10^6$ s (one month). We also present the results for the   first ALICE run at $2.76$ TeV. For the convenience of phenomenologists we provide here a parameterization of $\sigma_{\gamma\gamma}(\sqrt{s}=5.5\,\mathrm{TeV})$ as a function of the resonance mass. This makes simple the computation of event rates provided the specific meson  state and its two-photon decay width. We obtain in the interval $400\leq M_R\leq 4000$ MeV the parametrization:
\begin{eqnarray}
\frac{\sigma_{\mathrm{upc}}\, (AA\rightarrow R_J+AA)}{(2J+1)\,\Gamma(R_J\rightarrow \gamma\gamma)}= \frac{\sigma_0\,M^{\beta}_R}{1+\left(M_R/4\right)},
\end{eqnarray}
where $\sigma_0=4.9147$ mb/GeV and  $\beta=-3.45335$; $\Gamma_{\gamma\gamma}$ and $M_R$ are the decay width and  the resonance mass in units of GeV, respectively. The parameterization above allows to obtain estimates starting from a modeling for the two-photon width.

\begin{center}
\vskip 5pt
\begin{tabular}{|c||r|r|r|r|r|r|} \hline
Meson & \multicolumn{4}{c|}{LHC ($\sqrt{s} = 5.5$ TeV)} &
\multicolumn{2}{c|}{ALICE ($\sqrt{s} = 2.76$ TeV)} \\ \cline{2-7}
& $\Gamma_{\gamma\gamma}$ [eV] & $\Gamma_{gg}$ [MeV] & $\sigma_{\gamma\gamma}$ [$\mu$b] & $\sigma_{\pom\pom}$ [mb] &
$\sigma_{\gamma \gamma}$ [$\mu$b] & $\sigma_{\pom\pom}$ [mb]\\   \hline
$f_0(1500)$         & 0.77    &    69.8     &   1.30  & 1.07  & 0.78 & 0.99 \\
$f_0(1710)$         & 7.03    &    70.2    &   8.60   & 0.68  & 4.10 & 0.64 \\
$X(1835)$           & 0.02    &    70.3    & 0.02     & 0.54  & 0.01 & 0.50 \\
\hline
\end{tabular}
\end{center}
\vspace{-0.2cm}
Table 1: Integrated cross sections for channels $\gamma\gamma$ and $\pom\pom$ in $AA$ collisions at the LHC.

In Table 1 the results for Pomeron-Pomeron production of glueball is also presented. The estimates are shown for exclusive double Pomeron exchange, where estimations for sudakov suppression is disregarded (see discussion in Ref. \cite{Magno_PRC}). For the proton-proton case we obtained the estimates $\sigma_{\pom\pom}[f_0(1500)]= 116\,(128)$ $\mu$b,  $\sigma_{\pom\pom}[f_0(1710)]= 75\,(83)$ $\mu$b and $\sigma_{\pom\pom}[X(1835)]= 59\,(65)$ $\mu$b for energy of 7 TeV (14 TeV). The Pomeron contribution can be separated from photon channel by imposing a cut on
the impact parameter of collision. After imposing this kinematic cut ($b>2R_A$) the Pomeron contribution is reduced as they are dominated by small impact parameter contributions \cite{Schramm}.

As a final remark, it is important to discuss the uncertainties on the current calculations and the experimental feasibility of detecting glueballs candidates. The main uncertainty here is the model dependence on obtaining the two-photon and the two-gluon widths for a pure glueball meson. For the two-photon width we considered a nonrelativistic gluon bound-state model of Ref. \cite{Kada}, which it could be a debatable issue and it is far from being optimal. There are more modern approaches as reviewed in Ref. \cite{Crede}, but this is out of the scope of present work. For the two-gluon widths, we obtained them from the quarkonium width based on a non relativistic boundstate calculation \cite{close}. This type of matrix elements have been discussed in Refs. \cite{Chanowitz} giving rise to an effect of chiral suppression. We did not discuss the implication of those findings in present calculation. Concerning the experimental detection, the advantage of the exclusive processes discussed here is clear: glueballs are probably being produced with a high cross section in inelastic collisions (in $pp$ or $AA$ reactions) but when the multiplicity is high  the combinatorial background is overwhelming. In {\it exclusive} production there is no combinatorial background. In the ultraperipheral two-photon production of glueballs, the final state configuration is clear:  nuclei remain intact after collision and a double large rapidity gap between them is present (glueball is centrally produced with a low $p_T$ transverse momenta spectrum). This type of measurement is already done at RHIC for photoproduction of vector mesons and exclusive dilepton production with a  signal identification well understood \cite{Nystrand}. The situation for DPE glueball production is similar, with the $p_T$ spectrum being broader than the processes initiated by two-photons. Thus, a transverse momentum cut (and also impact parameter of collision) could separate the two channels.

\section*{Acknowledgements}
The author thanks V. Khoze, A. Szczurek and C. Pajares for their criticisms and useful remarks.

\end{document}